\documentclass[final,1p,times]{elsarticle}

\usepackage{amssymb}
\usepackage{hyperref}
\hypersetup{
    colorlinks = true,
}
\usepackage{url}
\usepackage{graphicx}
\usepackage{booktabs}
\usepackage{float}
\usepackage{enumitem}

\begin{document}

\begin{frontmatter}

\author[add1]{Pavlos Charalampidis}
\ead{pcharala@ics.forth.gr}
\author[add1]{Alexandros Fragkiadakis}
\ead{alfrag@ics.forth.gr}
\address[add1]{Institute of Computer Science\\
  Foundation for Research \& Technology - Hellas\\
  Heraklion, Crete, Greece}

\title{When Distributed Ledger Technology meets Internet of Things - Benefits and Challenges}

\begin{abstract}
There is a growing interest from both the academia and industry to employ distributed ledger technology in the Internet-of-Things domain for addressing security-related and performance challenges. Distributed ledger technology enables non-trusted entities to communicate and reach consensus in a fully distributed manner through a cryptographically secure and immutable ledger. However, significant challenges arise mainly related to transaction processing speed and user privacy. This work explores the interplay between Internet-of-Things and distributed ledger technology, analysing the fundamental characteristics of this technology and discussing the related benefits and challenges.

\end{abstract}


\end{frontmatter}

\section{Introduction}
Internet-of-Things (IoT) technologies and subsequent applications are remarkably increasing continuously, providing solutions in several areas like in industry, healthcare, agriculture, etc. This rapid proliferation has at the same time created challenges related to performance, security and privacy. As IoT networks are mainly based on severe resource constrained devices (sensors) in terms of memory, processing and storage, strong cryptographic primitives are difficult to apply. To defend against various cyber-attacks, a number of centralised trust-based schemes have been proposed (e.g.~\cite{fragkiadakis16}); however, their centralized nature make them highly vulnerable in case of single point failures (i.e when the central node responsible to detect attackers gets compromised or malfunctions). Other works, such as in~\cite{Feng11}, propose distributed schemes where each sensor monitors the behaviour of its neighbors and computes a trust value at the end of each round. Such schemes are vulnerable to various attacks, like Sybil attacks, as there are no countermeasures against identity theft.

IoT applications' operation could be significantly enhanced, if through suitable technological means, non-trusted peers (sensors, applications) can securely share information and reach consensus in a fully decentralized scheme. This can now become feasible using distributed ledger technology (DLT) that enables trustless communications using cryptographically secure immutable records. The most well-known type of DLT is the blockchain, the basis of the famous Bitcoin cryptocurrency~\cite{nakamoto2008bitcoin}. The use of DLT in the IoT domain can bring significant benefits, however not being a panacea, as various challenges also arise.

This article explores the interplay between IoT and DLT, discussing the benefits that DLT brings to the IoT ecosystem and highlights the challenges faced during the application of this technology, raising mainly due to the scale and the resource-constraint nature (in terms of end nodes) of the IoT networks. The rest of the paper is structured as follows. Section~\ref{sec:dlt} presents fundamental preliminaries of DLT. In Section~\ref{sec:consensus}, we describe various algorithms used for reaching consensus in a distributed network. DLT frameworks are presented in Section~\ref{sec:dlt_pl}. In Section~\ref{sec:iot}, we explore the interplay between IoT and DLT, while related applications are described in Section~\ref{sec:applications}. Finally, conclusions appear in Section~\ref{sec:conclusions}.

\section{Distributed ledger technology}
\label{sec:dlt}
The term distributed ledger refers to the general class of append-only databases that are maintained and shared among the mutually distrusting nodes of a network. Being broad as a term, in this article we consider DLT to be the combination of the following three components: (i) a peer-to-peer network, (ii) a distributed data storage, and (iii) cryptographic techniques for guaranteeing the validity of the ledger entries, its append-only nature and consensus in the network.

\subsection{Blockchain}
\begin{figure}
\centering
 \includegraphics[scale=0.5]{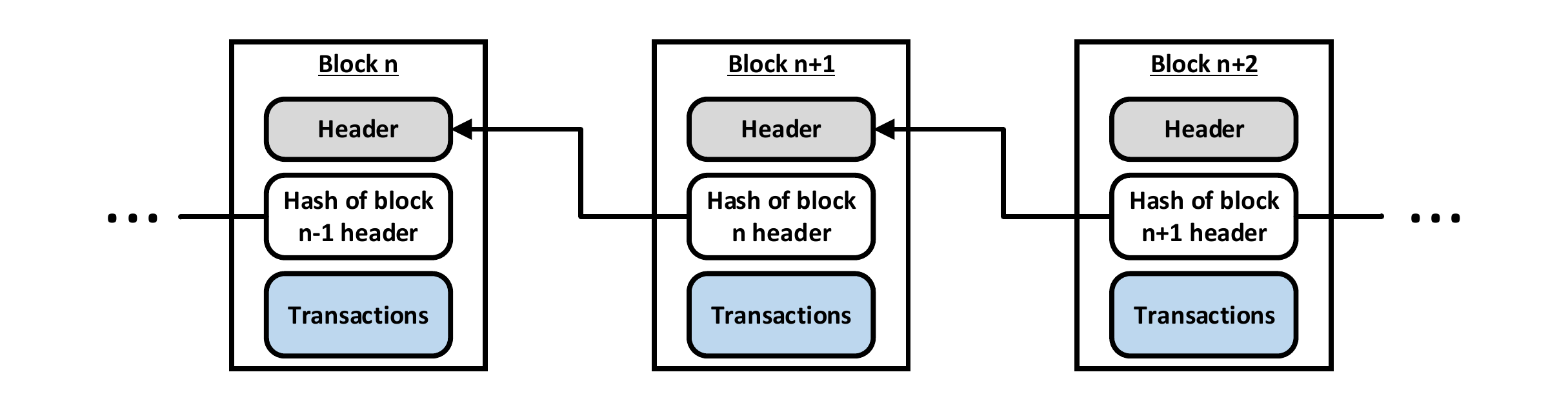}
  \caption{Block structure in a blockchain \label{fig:Fig1_blockchain}}
\end{figure}

The blockchain, being currently the most popular representative of DLT, is a distributed data structure that was introduced with Bitcoin cryptocurrency in order to solve the double spending problem. Essentially, it is a cryptographically secured and distributed data structure that acts as an immutable ledger of chronologically ordered blocks. In order to realize the chain in the blockchain, each block references the cryptographic hash of its preceding block (starting from the initial, the so-called \emph{genesis} block), thus creating a back-linked list, as shown in Figure~\ref{fig:Fig1_blockchain}. Including the hash of the previous block in the header of the next block, makes modifications to the blockchain almost impossible and ensures the append-only character of the ledger.

A blockchain is often maintained and populated by a distributed peer-to-peer network that, as stated before, comprises of mutually distrusting members. All members maintain a copy of the ledger and can interact following a decentralized paradigm that removes the need for a trusted intermediary. Thus, the blockchain network is said to support \emph{trustless communications}, in the sense that the network reaches consensus in terms of the transactional history without requiring parties to trust each other or any other central authority. By removing dependence on a centralized trusted authority, the blockchain avoids single point of decision (and possible failure) and accelerates the reconciliation between participating parties.

Each block in the data structure consists of a number of verified and mutually agreed transactions that describe the transfer of a \emph{digital tokenized asset}, such as cryptocurrency, between the nodes participating in the blockchain network. Asymmetric cryptography is leveraged for the realization of transactions. Each node is equipped with a private/public key pair; the private key is used for signing transactions of asset transfer while the public key provides node addressability in the network (each node is known by its public key or the hash of its public key).

Both \emph{validation} and \emph{consensus} on the transactions is essential due to the distrusting distributed environment that maintains the ledger. After each transaction is signed, it is broadcast to the network for verification against a set of conditions that are application-specific and must be met for any transaction to be permanently stored in the blockchain. Most commonly, this set of rules is hard-coded in the blockchain client that implements the protocol logic. In case a transaction is invalid, it is dropped, otherwise it is further broadcast, so that only valid transactions make it across the whole network. Transactions collected and validated during a predefined time interval, are packaged in a timestamped block by special nodes called \emph{miners}. As a last step, each newly-mined block is broadcast back to the network where it is validated for containing legit transactions and referencing through cryptographic hash its parent block.

Reaching consensus is of paramount importance for the viability of the network, otherwise the blockchain may exhibit \emph{forks}, i.e. inconsistent individual ledger copies. This can be seen as the core difference between a traditional database and a distributed ledger. Unfortunately, distributed consensus is traditionally known to suffer from scalability issues~\cite{vukolic2015quest}, nevertheless, a number of solutions has been proposed. For example, Bitcoin's protocol addresses the problem of decentralized consensus among a large number of nodes by adopting a probabilistic agreement approach, under the prerequisites of network synchronicity and legitimate behavior of the miners' majority. High scalability, however, comes at a severe cost in computational efficiency as well as transactions' throughput, dictating the selection of a consensus algorithm based on the application-dependent trade-off between performance and scalability. Consensus algorithms for distributed ledgers are further explored in Section~\ref{sec:consensus}.

Blockchain networks may adopt different authentication and authorization policy. Authentication policy defines which users have access to the blockchain. Thus, \emph{public} blockchains can be accessed by any user while \emph{private} blockchains are accessed only by users whitelisted by the blockchain owner. On the other hand, authorization policy determines what a user is authorized to do on the blockchain; in \emph{permissionless} blockchains any node can act as a miner/validator, in contrary to \emph{permissioned} blockchains where not all users are allowed to perform or validate transactions. Interestingly, different authorization policies are strongly related to different points in the centralization/decentralization spectrum. Thus, fully decentralized solutions (such as the Bitcoin and Ethereum network~\cite{buterin2013ethereum}) adopt a permissionless approach, where new users can freely join the network for validating transactions and mining blocks. By adopting a permissioned solution, the system generally shifts to a partially decentralized point, where transactions' creation and validation and/or block mining are orchestrated by a node oligarchy that is selected, e.g. by the network owner or the majority of the participants. Permissioned and permissionless blockchains exhibit also differences in terms of transaction processing rate, reversibility and finality, as well as flexibility in optimizing network rules~\cite{swanson2015consensus}.

\subsection{Smart contracts}
Smart contracts, a concept first introduced by Nick Szabo~\cite{szabo1997idea}, are abstractly described as computer programs that implement agreements between mutually distrusting parties, along with the mechanism for enforcing the agreement terms, without the existence of a trusted intermediary. In the DLT world, smart contracts are stored in the distributed ledger in an addressable form and their terms are automatically enforced by the consensus mechanism of the ledger. Smart contracts allow general purpose computations of complex logic, such as decision making through voting, crowdfunding and workflow management, safeguarded by the openness, immutability and verifiability offered by the DLT.

The complexity of the smart contract's logic is directly affected by the expressiveness of the scripting language a distributed ledger is built upon.
For example, the transactional model of Bitcoin, known as \emph{unspent transaction outputs (UTXO)} model (practically suitable for the trading/transfer of a digital asset between peers) utilizes a non-Turing complete scripting language in order to specify the conditions under which a transaction is redeemed. Nevertheless, expressiveness of Bitcoin's scripting language is bound to basic arithmetic, logical and cryptographic operations. On the other hand, the \emph{Ethereum} platform supporting the second largest cryptocurrency in terms of market capitalization, uses a stack-based bytecode language that is Turing-complete and enables the implementation of complex data-driven workflows (there is also a number of high level languages, such as Solidity~\cite{solidity} and Vyper~\cite{vyper}, which compile in Ethereum's bytecode). Ethereum is also diverging from the UTXO model of Bitcoin, by making smart contracts stored in the blockchain addressable by users or other smart contracts. This is known as the \emph{account model} and describes the ability of a contract to hold and exchange assets with users or other contracts, by receiving messages or transactions sent to its address, and maintain its internal state in a deterministic fashion.

Smart contracts have been recognized as the vehicle for the realization of the \emph{Distributed Autonomous Organization (DAO)} concept~\cite{chohan2017decentralized}. A DAO is described as a virtual entity that resides on the blockchain and whose behavior may be modified by invoking a certain process included in the DAO's smart contract. Behavior modification is possible through voting of the DAO's members, essentially a list of addresses that is carried by the smart contract. In reality, since the code of the smart contract is not mutable, a possible workaround is to let the behavior be implemented by different smart contracts whose addresses are known and modifiable by the DAO. Being responsible for handling assets of considerable value, it is of utmost importance that a smart contract's implementation is secured against asset stealing attacks, apart from consistent state representation and deterministic execution. In Ethereum network, successful attacks have been attributed to vulnerabilities related both to misunderstanding of Solidity's semantics as well as lack of a comprehensive and well-structured documentation~\cite{atzei2017survey}.

\subsection{Distributed ledgers based on Directed Acyclic Graphs}
Although blockchain is currently the most prevalent DLT paradigm, there is an effort for building distributed ledgers by leveraging \emph{Directed Acyclic Graphs (DAGs)}. The most notable difference between the two schemes is the way transactions are linked. On the one hand, blockchains bundle transactions in cryptographically linked blocks, which form a single back-linked list that encompasses the interactions' history, while, on the other hand, DAGs are built on transactional graphs, where each block or transaction represents a node in the graph. DAG distributed ledgers can be understood as a generalization of the back-linked blockchain that can improve transaction processing efficiency by allowing non-conflicting transactions from \emph{uncle blocks} (competing independently-mined blocks that originate from the same parent block) to be embodied in the main ledger. The looser validating rules for (blocks of) transactions in DAG-based DLT allow for relaxation of the network synchronization requirements. Thus, larger blocks with slower propagation time are also acceptable and transaction processing rate is considerably improved. DAG-related solutions for distributed ledgers come in two different flavors, namely \emph{blockDAG} and \emph{TDAG}~\cite{yeow2018decentralized}. In the first case, the DAG is constructed by blocks that reference more than one predecessors, while, in the second case, no blocks exist and the DAG is formed by separate transactions that reference the hashes of older transactions. TDAG solutions are shown to be more advantageous regarding confirmation speed and scalability compared to blockDAG ones.

Differences between blockchains and DAG-based DLT exist in terms of consensus mechanisms, transactions' confirmation, ledger size and scalability. For example, RaiBlocks implementation, recently known as Nano~\cite{lemahieu2018nano}, uses DAG to store transactions in a block-lattice structure. Transfer of asset is settled after both a ``send'' and ``receive'' transaction are performed. In terms of consensus, there is no need for leader election that will propose the next block of transactions to be added to the ledger (as is the case for Bitcoin or Ethereum). Instead, RaiBlocks employs a voting system of representatives in order to solve conflicts as well as vote for transactions' validity. Differently to blockchain frameworks, no dedicated nodes for transaction processing exist and the transactions' ordering is performed by each participant, posing no on-design bound to transaction processing throughput. That said, the limit is posed by network and hardware resources of the participants.

IOTA~\cite{popov2017tangle} is another DLT framework that is built on DAGs. It creates a TDAG, known as the \emph{Tangle}, that links transactions of tokenized assets. Each transaction also includes confirmation of two other transactions (by solving a computationally expensive hash puzzle), creating this way the DAG, with an edge pointing from each confirmed transaction to the newly created one. A weight proportional to the difficulty of the puzzle is assigned to each transaction that quantifies how strongly a transaction approves its predecessor transactions. This way, any new transaction, once released into the network, receives (at least partial) confirmations from peers almost instantly, removing the need for miners. As in Bitcoin, system stability depends on the legitimate behavior of the majority. IOTA Tangle was designed to support the IoT ecosystem, by being lightweight, highly scalable and omitting the necessity of fees for transaction processing, encouraging, thus, the micro-payments paradigm. In addition, it is claimed to be highly resistant to quantum computing attacks by employing hash-based one-time Winternitz signatures.

\section{Consensus algorithms}
\label{sec:consensus}
Distributed consensus in terms of the transactions' content and order included in the distributed ledger is of high importance for the network. This is a universal requirement irrespective of the DLT paradigm used and whose inability to satisfy may lead to inconsistencies in the copies of the ledger maintained by the network nodes. There is no single solution in the consensus problem and an algorithm is chosen based on the nature of the network, retaining otherwise the network's choices with regards to distribution, cryptographic immutability, as well as transparency.

The most important aspect in choosing a consensus algorithm is the access policy in network. In the case of a public and permissionless network, where anyone can enter, the consensus algorithm should be highly resistant to impersonation (Sybil) attacks. This is so, since a single entity could easily register with multiple identities, get into possession of multiple votes and thus control the network's decisions by pretending to have the majority. This constraint is clearly relaxed in the case of a permissioned network.

Permissionless networks typically try to mitigate Sybil attacks by making the voter's selection (in other words the mining process) computationally expensive. Consensus protocols for permissionless networks generally fall in the category of \emph{Proof-of-X} (PoX) mechanisms~\cite{tschorsch2016bitcoin}, with \emph{Proof-of-Work} (PoW) protocol being the most notable of them. In a PoW ledger a miner is permitted to append its candidate block only after having found the solution to a cryptographic hash-based puzzle of configurable difficulty. In the Bitcoin network a miner must calculate a partial hash collision, by finding a nonce in the block's header so that the $(SHA-256)^2$ hash of the header has an expected number of leading zeros. The puzzle is highly paralellizable and contemporary miners use specialized hardware, named as application-specific integrated circuits (ASICs), in order to win the voting race. As the computational power of miners increases, the difficulty of the hash puzzle also increases, so that the block generation rate remains constant and guaranties a loose synchronization of the network. However, the high computational resources required for solving a PoW puzzle require an economic incentive (usually in the form of a cryptocurrency) for motivating the miners to participate in the race. It is clear that the PoW protocol is highly inefficient, this inefficiency being the price for maintaining consensus in an open network.

\emph{Proof-of-Stake}~\cite{Li17} (PoS) is another popular consensus mechanism of the PoX family that seeks to address the inefficiencies of the PoW scheme. In a PoS supported network, the probability to win in the leader selection lottery is directly proportional to the participant's balance in terms of the asset being transacted (sometimes taking also in consideration the asset's age, e.g. in Peercoin's consensus mechanism~\cite{sunny2012ppcoin}). This approach removes the dependency of mining from computing power and depends it on deposit, eliminating high energy consumption. Participants holding a high asset balance are incentivized to act legitimately, since distrust to the network, caused by malicious activity, would affect negatively the value of the asset and, consequently, their (large) balances. A variant of PoS scheme, the \emph{Delegated Proof-of-Stake} (DPoS) mechanism~\cite{bach18dpos} shifts the PoS scheme closer to centralization in order to increase validation and transaction processing speed. To this end, a number of delegates is assigned the burden of configuring and performing block generation and validation. Delegates are voted by network participants, whose number of votes is proportional to their currency balance. In case of dishonest behavior, delegates are substituted. A limitation of the PoS mechanism that considers the stake age to determine leader selection probability, is that age is accumulated even when the node is offline. \emph{Proof-of-Activity} (PoA)~\cite{bentov2014proof} scheme solves this problem by rewarding nodes that participate actively to the network.

Both PoW and PoS strategies (and variants) are usually coupled with a cryptocurrency. A different PoX scheme that was firstly introduced in the Hyperledger Sawtooth platform for running general-purpose smart contracts on a distributed ledger on Intel hardware, is called as \emph{Proof-of-Elapse-Time} (PoET)~\cite{chen2017security}. It is based on the observation that PoW indirectly imposes a random waiting time for electing the leader that will append the next block to the ledger. For PoET consensus mechanism to be executed, the Intel Software Guard Extensions (SGX) hardware module is necessary. Each node in the network calls a secure container in SGX, known as \emph{enclave}, for setting a timer with a random delay that follows a probability distribution, which is determined by the scheme. The node whose timer is the first to expire becomes the leader that appends a new block. The system can create an \emph{attestation} that some particular trusted code was set up correctly in the protected environment and this attestation can be used by any node to verify that the leader has acted legitimately. This strategy removes the high power consumption required by PoW and can be also used in a permissioned distributed ledger network.

Contrary to permissionless blockchains, permissioned blockchain networks allow for the use of a wider scope of consensus mechanisms, since participating peers are whitelisted and the risk of impersonation attacks does not exist.

\emph{Practical Byzantine Fault Tolerance} (PBFT)~\cite{castro1999practical} is the archetypal algorithm used to solve the classical Byzantine Generals Problem for networks that are not synchronized. It requires at least two thirds of the peers to be legitimate. A leader is selected for proposing and ordering a transaction and its selection also requires approval of at least two thirds of the participants. In line with DPoS, there is also a delegated variant of PBFT, named as \emph{Delegated BFT} (DBFT), where voted delegates are responsible for generating and validating blocks.

Other variants of PBFT include the \emph{BFT-SMaRt}~\cite{sousa2013state} that is a high performance Byzantine Fault Tolerant State Machine Replication implementation. BFT-SMaRt is used in Symbiont~\cite{symbiont} and R3 Corda~\cite{r3corda} frameworks, as well as in Hyperledger Fablic v1.0~\cite{cachin2016architecture} where it is integrated as one of its ordering services. \emph{Tendermint Core}~\cite{kwon2014tendermint} uses also a variant of classic PBFT. Instead of sending a new transaction directly to all nodes, Tendermint clients disseminate their transactions to validate nodes through gossiping. Additionally, in contrast to PBFT, the protocol requires a constant rotation of the leader, which is changed after a block is appended to the ledger.

\emph{Ripple Protocol Consensus Algorithm} (RPCA)~\cite{schwartz2014ripple} uses a collectively trusted subnetworks approach for reaching consensus by introducing a \emph{convincing set} for each node. Each node creates a \emph{unique node list} (UNL), which includes the validating nodes that are sufficient for it to query for deciding whether to add a block to the ledger. In order to avoid forks, there must be a minimal overlap among the lists, at list one fifth of the longest list. The network can tolerate up to one fifth of faulty nodes in a UNL.

A dynamic permissioned model is adopted by \emph{MultiChain} platform~\cite{greenspan2015multichain}. A list of permitted nodes is responsible for validating blocks and participating in the protocol and this list may change through transactions that are executed on the blockchain. MultiChain uses the concept of \emph{mining diversity} to describe the number of blocks a miner should wait before trying to append a block to the chain. The protocol may suffer from forks the same way Bitcoin's PoW mechanism does, since two different nodes may produce a block almost at the same time. However, by following the ``longest-chain-wins'' rule, nodes finally converge to a single chain. Note that MultiChain consensus is characterized by lack of finality, just like PoW mechanism, being closer to the probabilistic agreement sense of consensus.

\section{Distributed ledger technology frameworks}
\label{sec:dlt_pl}
The landscape of frameworks for developing and managing DLT applications is highly fluid and platforms emanate from a variety of areas and sources that recognize the importance of DLT solutions. In this section, we present the most prominent frameworks, with an emphasis on those that seem to be most appropriate for developing IoT-related applications.

\subsection{Hyperledger}
Hyperledger~\cite{cachin2016architecture} is an umbrella project developed and maintained by the Linux Foundation with the goal of advancing cross-industry blockchain technologies. Instead of employing a single architecture, it is rather built on a modular and extensible basis, distinguishing and defining discrete components of a blockchain reference architecture. Most importantly, consensus and smart contracts functionality has been defined as two distinct layers; the \emph{consensus layer} is responsible for settling an agreement on the order and correctness of the transactions constituting a block, while the \emph{smart contract layer} implements request handling and validation of transactions executing business logic. Due to adoption of a permissioned approach (environment of partial trust), Hyperledger architecture encompasses a component for \emph{identity services} that is responsible for enrollment and registration of identities in the system, as well as authentication and authorization. Furthermore, there is a \emph{crypto abstraction} component that enables alternating between different cryptographic algorithms in a modular way.

In terms of consensus, both lottery-based and voting-based methods have been implemented, providing different guarantees for transaction processing speed, scalability and consensus finality. For example, Hypeledger Fabric is using Apache Kafka as the ordering service for fast finality and crash fault tolerance, but no Byzantine fault tolerance. Hypeledger Sawtooth, on the other hand, provides node scalability and Byzantine fault tolerance guaranteed by the PoET consensus mechanism. However, finality exists in the form of probabilistic agreement and can be delayed due to possible forks. Hypeledger Iroha provides Byzantine fault tolerance and quick finality, but requires that nodes are identified and fully connected. Interestingly, Hypeledger Fabric exhibits a broader understanding of the consensus concept that includes the whole transactional flow, that is transactional endorsement, ordering, and validation and commitment to the ledger. Consensus service is selectable and pluggable for all three phases, including various variants of BFT, such as BFT-SMaRt, Simple Byzantine Fault Tolerance (SBFT) etc.

Differences between separate frameworks also exist in the support and implementation of smart contracts. First of all, smart contracts may be either installed before network is launched or deployed on-chain as business logic that is stored on the ledger. State transition of the ledger is represented in different ways, such as change sets (Hypeledger Fabric), or cryptographically verifiable states (Hypeledger Sawtooth). In Hypeledger Fabric, smart contracts are programs that are named as \emph{chaincodes} and can be written in common programming languages, such as Go, JavaScript and Java. There are two chaincode types, namely \emph{system chaincode} and \emph{application chaincode}. The first one handles system-related transactions while the second one handles state of applications on the ledger. Smart contract layer stands in close cooperation with the consensus layer, since the consensus layer acts as a recommender on the contracts to be executed and verifies the identity of the entity requesting execution of a contract.

\subsection{Ethereum}
Ethereum~\cite{buterin2013ethereum} is a decentralized platform running on a global consensus-based virtual machine (Ethereum Virtual Machine) that is supported by a Turing-complete bytecode language for programming smart contracts. Primarily, it was developed for supporting \emph{Ether} cryptocurrency, however its flexible design permits execution of custom logic recorded on its blockchain. The blockchain operates on a permissionless mode (either public or private), which dictates the use of a mining process based on the highly inefficient PoW scheme, although there exist plans for switching to the PoS mechanism. Consensus is achieved at the ledger level, since it is crucial for the peers to agree on a definitive order of transactions and a global state to avoid double spends. Note that this stands in contrast with permissioned ledger of Hyperledger Fabric, where parties are only aware of the transactions they are involved into, thus consensus is practically reached at a transaction level. Ethereum lacks the modularity of Fabric (and generally Hyperledger), as regards consensus mechanisms, and pays a higher penalty in terms of performance scalability and transaction processing throughput. Having said that, the PoW scheme offers better scalability in terms of network size. In addition, it can easily enable implementation of decentralized applications that involve financial transactions, due to native support of cryptocurrency, and/or digital tokens' transfer, by constructing an appropriate smart contract.

Transactions and contracts to run on Ethereum require the consumption of an internal unit called \emph{gas}. Essentially, gas is the means for paying fees for transaction processing and contract execution, in a way that is proportional to the computational complexity of the process. Smart contract calculations are performed as long as there is still enough gas. This way the network is protected against wrongly written or malicious smart contracts that may cause unexpectedly long computations.

Ethereum smart contracts are generally written in a high level language, the most prominent of them being Solidity~\cite{solidity}. Solidity is a Javascript-like language, that adopts the so called \emph{contract-oriented} paradigm, which actually utilizes classes (as in object-oriented programming) to represent contracts. Each instance of the smart contract class is stored on the blockchain and is assigned an addressable account. State is stored in the smart contract's fields and methods are invoked by transactions addressing the smart contract. Apart from the common JavaScript semantics, Solidity encompasses a number of custom primitives for managing transaction and block state information. On the down side, Solidity's counter-intuitive semantics have been accused of causing a number of vulnerabilities in insecurely written smart contracts that have been involved in notorious Ethereum attacks~\cite{atzei2017survey} (e.g. DAO attack, King of the Ether Throne, etc.).

\subsection{Multichain}
Multichain platform ~\cite{greenspan2015multichain} is built on top of the Bitcoin's application programming interface (API), using a similar protocol format and transactional model. It extends the Bitcoin API for providing increased implementation flexibility and adding support for private blockchains used for storing digital assets of any kind. Essentially, Multichain tries to solve problems related to Bitcoin's PoW mining inefficiency, as well as privacy and openness of Bitcoin's blockchain, by integrating management of user permissions. Identity management and access control is performed by leveraging public key cryptography and practically relating permissions to a public address (mapping to a single public key) on the blockchain. The blockchain creator defines the initial set of administrators that are responsible for dynamically controlling permissions of users while the blockchain is active. Actions that require permission management are asset creation and exchange, as well as block generation. For permissions to alter, a predetermined proportion of administrators must reach consensus, with a possibly different proportion selected for each permission type.

Blockchains in Multichain platform are also flexible in terms of transaction and block size, as well as mining difficulty. In any case, in order to overcome the danger of mining monopoly in a permissioned private blockchain, the mining protocol relies on a randomized round-robin scheme for appending blocks and is controlled by a parameter known as \emph{mining diversity}. Mining diversity sets a constraint on the number of blocks the same miner is allowed to append within a given time window and, as a result, directly affects the percentage of miners that need to collude for attacking the network. A major limitation of the Multichain platform is the lack of support for smart contracts.

\subsection{Lisk}
Lisk platform~\cite{lisk} is based on sidechain technology that enables development of decentralized blockchain applications supported by different choices of cryptocurrency. The goal of Lisk is to enable users to write and deploy applications on a sidechain that further interacts with the main Lisk blockchain (named as \emph{mainchain} thus creating a universe of interoperable blockchains. Apart from transactions of the native cryptocurrency (known as LSK), custom digital assets can be also created and transferred in the mainchain. The platform is written in JavaScript and uses NodeJS and PostgreSQL for the backend.

Mining of new blocks is performed by utilizing the DPoS consensus scheme. The Lisk network is maintained by 101 delegates, each of whom is elected by the shareholders of LSK cryptocurrency. Delegates are ranked by the number of votes they receive and the top 101 of them are considered active, in the sense that only them are allowed to append a new block to the ledger. Consensus process is performed in rounds. At the beginning of each round, the relative order for block proposal is agreed between the delegates. Thus, in each round every delegate proposes one block that includes up to 25 transactions and, provided that the block is accepted by the system, the delegate earns a fixed number of LSKs. Lisk has been recently working with \emph{Chain of Things}~\cite{chainofthings} for addressing security issues within the IoT domain.

\section{Internet-of-Things and distributed ledger technology interplay}
\label{sec:iot}
The IoT paradigm is one of the most significant contemporary technologies that emerges as a natural evolution of the current Internet of computers. It comprises a large number of embedded cyber-physical devices (things) that have sensing and actuating capabilities, as well as the ability to exchange data with each other, forming a large interconnected ecosystem. The scale of the ecosystem that supports the development of decentralized and data-intensive applications running on billions of devices, necessitates addressing interoperability issues between IoT systems, or else vertical silos developed may hinder IoT reaching its full potential. Interoperability, however, demands trust between interacting platforms. In addition, the pervasiveness and invisibility of data collection, the heterogeneity of devices along with the scale of an IoT system raise several security and privacy issues.

\subsection{Distributed ledger technology advantages}

DLT has been recently proposed as a tool for addressing security and privacy challenges, since they encompass key features that can be advantageous for the IoT domain (Table~\ref{tab:DLTsolutions}). The ever expanding IoT ecosystem necessitates solutions of decentralized and distributed nature. Decentralization offers scalability and robustness that facilitates maintenance and management and avoids the pitfall of single point of decision and, thus, possible failure. In addition, it removes bottlenecks in the information flow, as well as its control by an oligarchy of powerful stakeholders. DLT is decentralized by design, since it is supported by a peer-to-peer network for reaching consensus, and achieves resilience and high availability by employing multiple copies of the ledger. The architecture of DLT is able to mitigate connectivity or service provision failures that could be catastrophic for several IoT applications.

\begin{table}[ht]
	\resizebox{\textwidth}{!}{%
    \begin{tabular}{p{.4\textwidth} p{.6\textwidth}}
        \toprule IoT challenges & Solutions offered by DLT\\\midrule
         Exponential growth in number of IoT devices & Scalable architecture that facilitates network management and avoids single points of failure\\\\
         Centralized information flow & Decentralization by design that offers transparency in a trustless environment\\\\
         Broad attack surfaces in terms of security and privacy & Autonomous identity management, access control and auditability\\\\
         Data sources originating from private verticals & Device-to-device marketplaces supported by native cryptocurrencies\\\\
        \bottomrule
    \end{tabular}
    }
    \caption{Solutions offered by DLT to IoT ecosystem}  \label{tab:DLTsolutions}
\end{table}

The limited capabilities of the IoT devices in terms of processing, memory and storage, in addition to their autonomous operation and communication make them highly vulnerable to security threats. High decentralization and autonomous features make DLT a strong candidate for securing IoT systems, since they can offer security through transparency in a trustless IoT environment. DLT can provide a means for securely configuring and controlling operation modes of IoT devices, by preventing unauthorized access. In addition, DLT enables secure validation of the identity of an IoT device. The public key of an asymmetric key pair can be stored in the tamper-resistant blockchain, while the corresponding private key can be stored in an embedded cryptographic chip on the device. Any network node is able to verify the identity of a device, by accessing its public key that lies in the blockchain and challenging the device, offering in this way secure authentication and protection against identity spoofing. Finally, DLT can be leveraged for implementing access control in IoT systems that shifts from the fail-prone and inefficient centralized model and adopts a decentralized approach, which establishes policy enforcement points on the edge IoT devices and policy decision points on the system validators/miners. Appropriate smart contracts residing on the ledger can be developed for enforcing the access policies in an autonomous and verifiable manner.

Applications empowered by IoT commonly require autonomous operation of the devices, in order to support solutions such as Machine-as-a-Service or smart energy markets. DLT enables machine-to-machine interaction without the involvement of servers, so that applications can offer testbed-agnostic and device-decoupled services. In addition, DLT provides auditability out-of-the-box, so that it is possible to track the network's behavior in a verifiable and tamper-resistant manner, by taking advantage of the digital timestamping feature of the ledger. Auditability improves data analytics, network performance and legal compliance, as well as provides a strong tool in dispute resolution between disagreeing peers.

Data generated by IoT devices is mostly owned by device owners and is commonly private by nature. However, this situation can hinder unraveling the full potential of the IoT ecosystem by creating impermeable verticals. An opportunity raises to create a marketplace where data producing devices can sell data to consuming devices, e.g. following a Sensing-as-a-Service paradigm~\cite{perera2014sensing}. By further extending the idea, IoT devices can provide services (e.g. disk space, API calls etc.) for receiving remuneration. A DLT-based cryptocurrency (supporting micro-payments and being even feeless) specifically tailored for the heterogeneous and resource-constraint nature of IoT can act as the billing layer for such a marketplace~\cite{christidis2016blockchains}.

\subsection{Challenges in the IoT domain}

Although DLT presents itself as a promising means of strengthening the security, privacy and trustworthiness of the IoT ecosystem, the participation of IoT devices in a DLT network comes with a number of challenges that should be addressed for developing viable and robust applications.

Trustless decentralization and reconciliation come at a cost in terms of \emph{transactional processing speed}, especially in the case of public networks, where the risk of impersonation attacks dictates the use of expensive mechanisms for block mining. Even if the maximum number of transactions per block is increased so as to achieve a higher throughput, this would come with a cost to transactions' validation time. The challenge is even more profound in the case of smart contracts, where parallelization of tasks is not possible due to concurrency issues. In addition, the resource-constrained nature of most IoT devices stands in stark contrast to blockchain requirements in terms of computations, storage and network bandwidth. It is clear that low class IoT devices would face severe difficulties in maintaining a blockchain, so hybrid architectures (where the computational bottleneck is shifted to more powerful devices, e.g. gateways) or multiple local/multi-layer blockchains should be considered as a solution. This strategy, however, implies that mining power would be the privilege of a smaller subset of the network nodes, increasing the risk of successful collusions and unfair transaction handling. Using computationally cheaper consensus mechanisms, such as PoET or PBFT, can be also a line of action. In addition, cheaper hashing algorithms, such as Scrypt (a memory intensive password-based key derivation function), could be used in order to decrease mining energy cost. On the other hand, increased communication and storage requirements can be reduced through the use of sidechains~\cite{back2014enabling} (to partially cut-through transactions and drop requirements of full history verification) or mini-blockchains~\cite{bruce2014mini}, where old transactions are pruned, in order to decrease computational burden of full nodes, while blockchain headers are kept to maintain the ability to verify the longest blockchain.

Although not IoT-specific, challenges in terms of privacy of participants, and confidentiality and integrity of the transactions, are existent and extremely significant. Privacy issues are mainly related to the preservation of participants' anonymity. In a public blockchain all transactions are transparent and prone to analysis by an interested party, thus privacy should be maintained by ``breaking the flow of information'', so that it is difficult to identify patterns and link addresses to real identities. A number of studies~\cite{koshy2014analysis} \cite{feld2014analyzing} has shown evidence of anonymity reduction by exploiting traffic analysis tools that can e.g. map Bitcoin addresses to IP addresses or link transactions by observing that an average peer-list contains addresses that mostly reside in the own autonomous systems of the peers. Mitigation of anonymity issues has been achieved by mixing protocols~\cite{ziegeldorf2015coinparty} that enable coins' transfer between user addresses without a clear trace linking between the addresses. It is clear that this strategy works in case of cryptocurrency DLT. In a more general scenario, irrelevant to the asset transacted, the use of hierarchical deterministic wallets~\cite{hdw2018} allows the safe and controlled generation of an infinite number of public keys. Thus, a fresh key can be used for every transaction, strengthening this way the anonymity of the network.

Confidentiality of the transactions is also difficult to obtain, since transactions are stored publicly in the distributed ledger (e.g. transactions sent and stored in a smart contract). One potential solution comes in the form of a zero-knowledge proofs protocol~\cite{ben2013snarks} (e.g. \emph{zk-SNARKs}) that enables validating the truth of a statement without revealing its content. As a result, in a zero-knowledge transaction, the network peers are able to validate a transaction that has taken place, but learn nothing about the sender, the recipient or the transferred asset and its quantity. Another approach would be to exploit the \emph{state channels paradigm}, where participants exchange several cryptographically signed messages, accumulating intermediate state changes, without publishing them to the canonical chain until the channel is closed. A reference is then recorded to the blockchain, so as to avoid posting intermediate sensitive transactions. This can be of high importance in fields where the competitive advantage can be lost, in case information stored in the ledger is leaked to competitors, e.g. in industrial or financial domain.

Transactions' integrity concerns raise in relation to miners' faulty or malicious behavior that can lead to the creation of blockchain forks. The PoW mechanism is designed with the assumption that honest miners control the network. If malicious miners collectively control more computational power than the legitimate ones, the network is vulnerable to the so called 51\% attack. A \emph{selfish mining attack}~\cite{eyal2018majority} is also known, where colluding miners obtain a revenue larger than their fair share by keeping discovered blocks private, thus building a private fork. The fork is published as soon as it catches up with the length of the public blockchain and has chances of becoming the longest one in the network by being accepted by the legitimate nodes. Delay in the delivery of blocks or transactions by exploiting the scalability measures implemented in the Bitcoin blockchain are also possible, enabling DoS scenarios (since a powerful attacker can in such way prevent data dissemination in the network) and facilitating selfish mining by avoid propagating mined blocks. A summary of challenges faced in the application of DLT solutions to the IoT domain is presented in Table~\ref{tab:DLTchallenges}.

\begin{table}[ht]
	\resizebox{\textwidth}{!}{%
    \begin{tabular}{p{.3\textwidth} p{.8\textwidth}}
        \toprule Challenges & Countermeasures\\\midrule
        Transaction processing speed & \vspace{-\topsep} \begin{itemize}[leftmargin=*]
            \item Adopt less expensive consensus algorithms (e.g. PBFT or PoET over PoW)
            \item Use faster hashing algorithms, e.g. Scrypt
            \item Use of sidechains/mini-blockchains for pruning transactions and reduce verification requirements
        \end{itemize}\\

         User anonymity & \vspace{-\topsep} \begin{itemize}[leftmargin=*]
            \item Use mixing protocols
            \item Adopt hierarchical deterministic wallets for key renewal
        \end{itemize}\\

        Storage capacity & \vspace{-\topsep} \begin{itemize}[leftmargin=*]
            \item Store large data off chain/store data hash instead of full data
			\item Use distributed file sharing systems, e.g. Inter Planetary File System (IPFS)
        \end{itemize}\\

		Data privacy & \vspace{-\topsep} \begin{itemize}[leftmargin=*]
            \item Impose access control/user permissions on transactions
			\item Offload sensitive data to sidechains
			\item Use homomorphic encryption/zero-knowledge proofs
        \end{itemize}\\

        \bottomrule
    \end{tabular}
    }
    \caption{Challenges in DLT-supported IoT deployments}  \label{tab:DLTchallenges}
\end{table}

\section{DLT applications in the IoT}
\label{sec:applications}

Only recently both academia and industry have begun to experiment with the integration of DLT solutions for improving the IoT technology in terms of decentralization, security, reliability and autonomous operation. Nevertheless, a significant number of applications has emerged as a result of this effort. In this section we present the most prominent applications that highlight the positive impact DLT can have on IoT systems.

\subsection{Software updates}
\begin{figure}
\centering
 \includegraphics[scale=0.7]{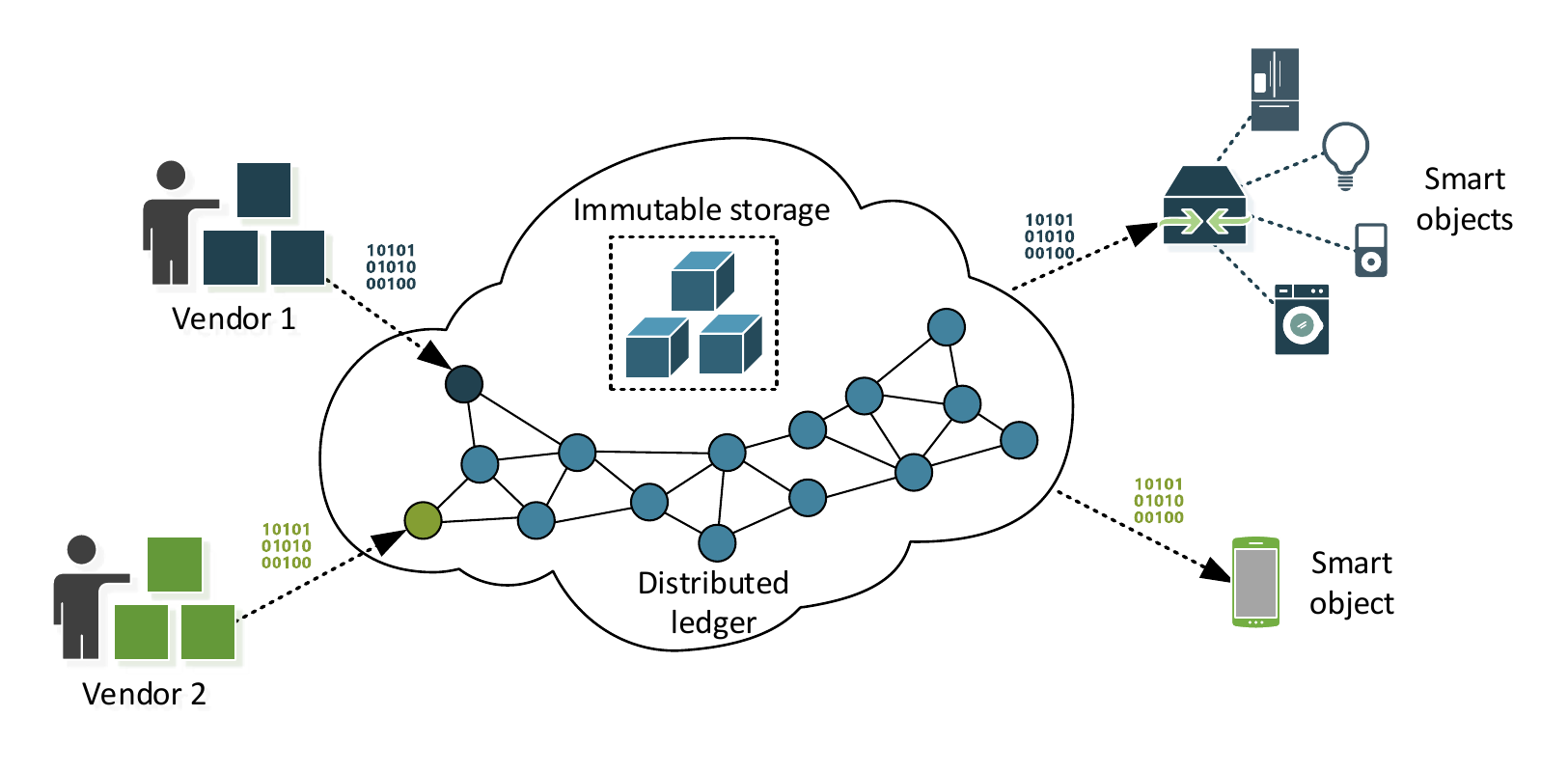}
  \caption{DLT-assisted IoT software update  \label{fig:Fig2_software_update}}
\end{figure}

Limited capabilities and resources of IoT devices make the use of strong and recognized security mechanisms difficult to apply. As a result, vulnerabilities in the software or firmware of the devices are a recurring issue that needs emergent response in order to avoid exploitation of the devices by malicious attackers. Nevertheless, the large scale of IoT networks makes centralized software distribution challenging and costly. Blockchain technology in cooperation with a distributed peer-to-peer file system that plays the role of an immutable storage (e.g. IPFS~\cite{benet2014ipfs}, Swarm~\cite{hartman1999swarm}) can offer a means of automatic distribution of the most recent software binary in a verifiable and highly robust (due to the large number of the binary's replicas) way, as shown in Figure~\ref{fig:Fig2_software_update}. A protocol for secure firmware update of embedded IoT devices is proposed in~\cite{lee2017blockchain}. The hash of the latest firmware is stored in the blockchain, so that an embedded device can query it for examining the validity and the freshness of the firmware version it is currently running. If validity test fails or a new version of the firmware has been released, the device can retrieve the latest valid firmware through a BitTorrent based peer-to-peer sharing network.

\subsection{Supply chain}
IoT-assisted supply chain management can be benefited by the adoption of a blockchain solution. Instead of using an internal database for each stakeholder being part of the supply-chain, a blockchain network can track a specific asset by exposing a shared verifiable database that can be automatically propagated along the network. Thus, the information can be trailed in an auditable manner. The process can be implemented either by following the UTXO model of atomic digital asset exchanges or by addressing signed messages to a smart contract that resides on the blockchain (e.g. Ethereum blockchain), whenever the asset changes hands from one stakeholder to the other, so as to prove satisfaction of contractual obligations between the shareholders~\cite{christidis2016blockchains}.

\subsection{Smart home}
A plethora of household devices related to environmental control, household electronics and home security systems are present in a modern smart home. Home owners can monitor and control the installed IoT devices remotely, while information from sensing devices is usually stored in a centralized server (or gateway) which provides access to the authorized user for information collection or home control. In order to eliminate the single point of failure due to the centralized access point, a smart home system can be benefited by a blockchain approach that can store in a verifiable manner digests of device communications, sensory data and owner access requests. This is information critical both for resilience and anomaly detection and logging. In~\cite{dorri2017smarthome} the authors propose a lightweight blockchain-based architecture for smart home management. The architecture is constituted by three tiers, namely cloud storage, overlay and smart home. Each home possesses a home miner (a high-resource device) that maintains a local private blockchain. Transactions performed by home devices are stored in the local blockchain in blocks that are mined without any hashing puzzles, in order to improve transactional processing efficiency. On the second layer, the home miners form a clustered overlay peer-to-peer network that maintains a public overlay blockchain and stores transactions sent by the cloud storage as well as owner access transactions. No cryptographic puzzles are used for mining blocks in the overlay blockchain, and a complementary distributed trust management mechanism controls the trust of clusterhead-miners. A qualitative analysis demonstrates how the system achieves confidentiality, integrity and availability and prevents attacks, such as Denial-of-Service or linking attack.

\subsection{Automotive security}
Wireless vehicle interface of future smart vehicles will enable communication and data exchange with other vehicles or IoT devices, on behalf of the driver, for the provision of advanced services, such as traffic management, accident response, car problem diagnosis, and integrated navigation. Data of sensitive nature are commonly exchanged, necessitating the addressing of several security and privacy challenges, in addition to scalability and maintainability issues. Blockchain offers an elegant way to achieve decentralization and tamper-proof logging of the vehicle communications and car-related data, while also ensuring security and privacy~\cite{dorri2017automotive}. Benefits can be foreseen in applications, such as over-the-air software updates for smart vehicles (akin to the first use case described here) or flexible insurance fees based on driver behavior.

\subsection{Smart energy markets}
Energy markets that enable entities acting simultaneously as consumers and producers (aka \emph{prosumers}) of electrical energy from renewable energy sources, balance local generation and consumption in a decentralized fashion. A DLT peer-to-peer network may provide a facilitating substrate for automatic energy trading by autonomous agents, based on a predefined set of rules (expressed through a smart contract), without the intervention of centralized trusted intermediaries. DLT enables fully decentralized market platforms by resolving conflicts and providing democratic and symmetric participation to all members, as well as providing a transparent and secure log for tracing transactions of energy. Authors in~\cite{mengelkamp2018microgrid} present a microgrid double auction market for renewable (solar) energy trading (localized in Brooklyn, NY) that is implemented and operated by a private blockchain based on Tendermint  protocol. The same authors provide in~\cite{mengelkamp2017smartgrid} a preliminary economic evaluation of their local energy market platform by showing the potential electricity cost reduction for its participants.

\subsection{Smart healthcare}
IoT systems have been proven suitable for applications related to medical care, by interconnecting available medical resources and providing smart and reliable healthcare services. Interplay between DLT and IoT can benefit the healthcare domain, for instance, in pharmaceutical supply chain, clinical trials and precision medicine. In~\cite{bocek2017blockchains}, the authors design and implement an architecture for monitoring temperature conditions during the transport of medical products. Parcels are equipped with temperature and humidity sensors that transfer their collected data to Ethereum blockchain. A smart contract that lies on the blockchain determines the quality of the medical product by examining if the reported values remain within the accepted range. Sensor data integrity and accessibility is guaranteed by the tamper-proof and public nature of the blockchain. In~\cite{salahuddin2018softwarization}, an IoT architecture for secure smart healthcare is proposed that achieves dynamic configurability by utilizing Software Defined Networking (SDN) and Network Functions Virtualization (NFV). A blockchain solution is envisioned for guaranteeing security of sensitive data by tracking and protecting against unauthorized access to confidential medical records.

\section{Conclusions}
\label{sec:conclusions}
In this work, we explored the interplay between IoT and DLT, showing that security and performance-related challenges can be addressed, benefiting from cryptographically secure immutable ledgers. Non-trusted IoT entities (applications, sensors, etc.) are able to reach consensus over trustless communication. Significant challenges arise such as the transaction processing speed, user privacy, storage and data capacity, etc. Nevertheless, the presented applications and DLT platforms show that there is an increasing interest from both the academia and industry to advance DLT in the IoT domain.

\section*{Acknowledgements}
{This research has been financed by the European Union and Greek national funds through the Operational Program Competitiveness, Entrepreneurship and Innovation, under the call RESEARCH – CREATE – INNOVATE (project code: T1EDK-00070).}

\bibliographystyle{elsarticle-num}
\bibliography{refs}

\begin{thebibliography}{10}
\expandafter\ifx\csname url\endcsname\relax
  \def\url#1{\texttt{#1}}\fi
\expandafter\ifx\csname urlprefix\endcsname\relax\def\urlprefix{URL }\fi
\expandafter\ifx\csname href\endcsname\relax
  \def\href#1#2{#2} \def\path#1{#1}\fi

\bibitem{fragkiadakis16}
A.~Fragkiadakis, E.~Tragos, {A Trust-Based Scheme Employing Evidence Reasoning
  for IoT Architectures}, in: {WF-IoT}, 2016, pp. 559--564.

\bibitem{Feng11}
R.~Feng, X.~Xu, X.~Zhou, J.~Wan, {A Trust Evaluation Algorithm for Wireless
  Sensor Networks Based on Node Behaviors and D-S Evidence Theory}, {Sensors}
  11 (2011) 1345--1360.

\bibitem{nakamoto2008bitcoin}
S.~Nakamoto, {Bitcoin: A peer-to-peer electronic cash system},
  \url{https://bitcoin.org/bitcoin.pdf} (2008).

\bibitem{vukolic2015quest}
M.~Vukoli{\'c}, {The quest for scalable blockchain fabric: Proof-of-work vs.
  BFT replication}, in: International Workshop on Open Problems in Network
  Security, Springer, 2015, pp. 112--125.

\bibitem{buterin2013ethereum}
V.~Buterin, Ethereum white paper, \url{https://ethereum.org/en/whitepaper/}
  (2013).

\bibitem{swanson2015consensus}
T.~Swanson, {Consensus-as-a-service: a brief report on the emergence of
  permissioned, distributed ledger systems},
  \url{http://www.ofnumbers.com/wp-content/uploads/2015/04/Permissioned-distributed-ledgers.pdf}
  (2015).

\bibitem{szabo1997idea}
N.~Szabo, The idea of smart contracts, Nick Szabo's Papers and Concise
  Tutorials 6 (1997).

\bibitem{solidity}
Solidity, \url{https://solidity.readthedocs.io/}.

\bibitem{vyper}
Vyper, \url{https://vyper.readthedocs.io/}.

\bibitem{chohan2017decentralized}
U.~W. Chohan, The decentralized autonomous organization and governance issues,
  Available at SSRN 3082055 (2017).

\bibitem{atzei2017survey}
N.~Atzei, M.~Bartoletti, T.~Cimoli, {A survey of attacks on Ethereum smart
  contracts}, in: Internation Conference on Principles of Security and Trust,
  Springer, 2017, pp. 164--186.

\bibitem{yeow2018decentralized}
K.~Yeow, A.~Gani, R.~W. Ahmad, J.~J. Rodrigues, K.~Ko, Decentralized consensus
  for edge-centric internet of things: A review, taxonomy, and research issues,
  IEEE Access 6 (2018) 1513--1524.

\bibitem{lemahieu2018nano}
C.~LeMahieu, {Nano: A Feeless Distributed Cryptocurrency Network},
  \url{https://nano.org/en/whitepaper} (2018).

\bibitem{popov2017tangle}
S.~Popov, {The Tangle},
  \url{https://assets.ctfassets.net/r1dr6vzfxhev/2t4uxvsIqk0EUau6g2sw0g/45eae33637ca92f85dd9f4a3a218e1ec/iota1_4_3.pdf}
  (2018).

\bibitem{tschorsch2016bitcoin}
F.~Tschorsch, B.~Scheuermann, Bitcoin and beyond: A technical survey on
  decentralized digital currencies, IEEE Communications Surveys \& Tutorials
  18~(3) (2016) 2084--2123.

\bibitem{Li17}
W.~Li, S.~Andreina, J.-M. Bohli, G.~Karame, {Securing Proof-of-Stake Blockchain
  Protocols}, in: Data Privacy Management, Cryptocurrencies and Blockchain
  Technology, Springer, 2017, pp. 297--315.

\bibitem{sunny2012ppcoin}
S.~King, S.~Nadal, {PPCoin: Peer-to-Peer Crypto-Currency with Proof-of-Stake},
  \url{https://peercoin.net/assets/paper/peercoin-paper.pdf} (2012).

\bibitem{bach18dpos}
L.~M. {Bach}, B.~{Mihaljevic}, M.~{Zagar}, Comparative analysis of blockchain
  consensus algorithms, in: 2018 41st International Convention on Information
  and Communication Technology, Electronics and Microelectronics (MIPRO), 2018,
  pp. 1545--1550.

\bibitem{bentov2014proof}
I.~Bentov, C.~Lee, A.~Mizrahi, M.~Rosenfeld, {Proof of Activity: Extending
  Bitcoin’s Proof of Work via Proof of Stake}, ACM SIGMETRICS Performance
  Evaluation Review 42~(3) (2014) 34--37.

\bibitem{chen2017security}
L.~Chen, L.~Xu, N.~Shah, Z.~Gao, Y.~Lu, W.~Shi, {On Security Analysis of
  Proof-of-Elapsed-Time (PoET)}, in: International Symposium on Stabilization,
  Safety, and Security of Distributed Systems, Springer, 2017, pp. 282--297.

\bibitem{castro1999practical}
M.~Castro, B.~Liskov, et~al., {Practical Byzantine Fault Tolerance}, in: {3rd
  OSDI (Operating Systems Design and Implementation)}, Vol.~99, 1999, pp.
  173--186.

\bibitem{sousa2013state}
J.~Sousa, E.~Alchieri, A.~Bessani, {State machine replication for the masses
  with BFT-SMART}, in: {44th Annual IEEE/IFIP International Conference on
  Dependable Systems and Network}, 2014, pp. 355--362.

\bibitem{symbiont}
Symbiont, \url{https://symbiont.io/}.

\bibitem{r3corda}
Corda, \url{https://www.corda.net/}.

\bibitem{cachin2016architecture}
C.~Cachin, Architecture of the hyperledger blockchain fabric, in: {Workshop on
  Distributed Cryp-tocurrencies and Consensus Ledgers (DCCL 2016)}, 2016.

\bibitem{kwon2014tendermint}
J.~Kwon, {Tendermint: Consensus without Mining},
  \url{https://tendermint.com/static/docs/tendermint.pdf} (2014).

\bibitem{schwartz2014ripple}
D.~Schwartz, N.~Youngs, A.~Britto, et~al., {The Ripple Protocol Consensus
  Algorithm}, \url{https://ripple.com/files/ripple_consensus_whitepaper.pdf}
  (2014).

\bibitem{greenspan2015multichain}
G.~Greenspan, {MultiChain Private Blockchain - White paper},
  \url{http://www.multichain.com/download/MultiChain-White-Paper.pdf} (2015).

\bibitem{lisk}
Lisk, \url{https://lisk.io/}.

\bibitem{chainofthings}
{Chain of Things}, \url{https://www.chainofthings.com/}.

\bibitem{perera2014sensing}
C.~Perera, A.~Zaslavsky, P.~Christen, D.~Georgakopoulos, {Sensing as a Service
  Model for Smart Cities Supported by Internet of Things}, Transactions on
  Emerging Telecommunications Technologies 25~(1) (2014) 81--93.

\bibitem{christidis2016blockchains}
K.~Christidis, M.~Devetsikiotis, {Blockchains and smart contracts for the
  Internet of Things}, {IEEE Access} 4 (2016) 2292--2303.

\bibitem{back2014enabling}
A.~Back, M.~Corallo, L.~Dashjr, M.~Friedenbach, G.~Maxwell, A.~Miller,
  A.~Poelstra, J.~Tim{\'o}n, P.~Wuille, Enabling blockchain innovations with
  pegged sidechains, \url{https://blockstream.com/sidechains.pdf} (2014).

\bibitem{bruce2014mini}
J.~Bruce, {The Mini-Blockchain Scheme},
  \url{http://cryptonite.info/files/mbc-scheme-rev3.pdf} (2017).

\bibitem{koshy2014analysis}
P.~Koshy, D.~Koshy, P.~McDaniel, An analysis of anonymity in bitcoin using p2p
  network traffic, in: International Conference on Financial Cryptography and
  Data Security, Springer, 2014, pp. 469--485.

\bibitem{feld2014analyzing}
S.~Feld, M.~Sch{\"o}nfeld, M.~Werner, {Analyzing the Deployment of Bitcoin's
  P2P Network under an AS-level Perspective}, Procedia Computer Science 32
  (2014) 1121--1126.

\bibitem{ziegeldorf2015coinparty}
J.~H. Ziegeldorf, F.~Grossmann, M.~Henze, N.~Inden, K.~Wehrle, {Coinparty:
  Secure multi-party mixing of bitcoins}, in: Proceedings of the 5th ACM
  Conference on Data and Application Security and Privacy, ACM, 2015, pp.
  75--86.

\bibitem{hdw2018}
Bip0032: Hierarchical deterministic wallets,
  \url{https://github.com/bitcoin/bips/blob/master/bip-0032.mediawiki}.

\bibitem{ben2013snarks}
E.~Ben-Sasson, A.~Chiesa, D.~Genkin, E.~Tromer, M.~Virza, {SNARKs for C:
  Verifying program executions succinctly and in zero knowledge}, in: Advances
  in Cryptology--CRYPTO 2013, Springer, 2013, pp. 90--108.

\bibitem{eyal2018majority}
I.~Eyal, E.~G. Sirer, {Majority is not enough: Bitcoin mining is vulnerable},
  Communications of the ACM 61~(7) (2018) 95--102.

\bibitem{benet2014ipfs}
J.~Benet, Ipfs - content addressed, versioned, p2p file system, ArXiv
  abs/1407.3561 (2014).

\bibitem{hartman1999swarm}
J.~H. Hartman, I.~Murdock, T.~Spalink, {The Swarm scalable storage system}, in:
  Distributed Computing Systems, 1999. Proceedings. 19th IEEE International
  Conference on, IEEE, 1999, pp. 74--81.

\bibitem{lee2017blockchain}
B.~Lee, J.-H. Lee, {Blockchain-based secure firmware update for embedded
  devices in an Internet of Things environment}, The Journal of Supercomputing
  73~(3) (2017) 1152--1167.

\bibitem{dorri2017smarthome}
A.~Dorri, S.~S. Kanhere, R.~Jurdak, P.~Gauravaram, {Blockchain for IoT security
  and privacy: The case study of a smart home}, in: Pervasive Computing and
  Communications Workshops (PerCom Workshops), 2017 IEEE International
  Conference on, IEEE, 2017, pp. 618--623.

\bibitem{dorri2017automotive}
A.~Dorri, M.~Steger, S.~S. Kanhere, R.~Jurdak, Blockchain: A distributed
  solution to automotive security and privacy, IEEE Communications Magazine
  55~(12) (2017) 119--125.

\bibitem{mengelkamp2018microgrid}
E.~Mengelkamp, J.~G{\"a}rttner, K.~Rock, S.~Kessler, L.~Orsini, C.~Weinhardt,
  {Designing microgrid energy markets: A case study: The Brooklyn Microgrid},
  Applied Energy 210 (2018) 870--880.

\bibitem{mengelkamp2017smartgrid}
E.~Mengelkamp, B.~Notheisen, C.~Beer, D.~Dauer, C.~Weinhardt, A
  blockchain-based smart grid: towards sustainable local energy markets,
  Computer Science-Research and Development (2017) 1--8.

\bibitem{bocek2017blockchains}
T.~Bocek, B.~B. Rodrigues, T.~Strasser, B.~Stiller, Blockchains everywhere-a
  use-case of blockchains in the pharma supply-chain, in: Integrated Network
  and Service Management (IM), 2017 IFIP/IEEE Symposium on, IEEE, 2017, pp.
  772--777.

\bibitem{salahuddin2018softwarization}
M.~A. Salahuddin, A.~Al-Fuqaha, M.~Guizani, K.~Shuaib, F.~Sallabi,
  {Softwarization of Internet of Things Infrastructure for Secure and Smart
  Healthcare}, Computer 50~(7) (2017) 74--79.

\end{thebibliography}

\end{document}